\documentclass[%
reprint,
superscriptaddress,
 amsmath,amssymb,
 aps,
floatfix,
]{revtex4-1}

\usepackage{color}

\usepackage{graphicx}
\usepackage{dcolumn}
\usepackage{bm}
\usepackage{mathrsfs}
\usepackage[normalem]{ulem}

\begin{document}

\preprint{APS/123-QED}
\title{Origin of the asymmetric light emission from molecular exciton-polaritons}

\author{Tom\'{a}\v{s} Neuman}
\
\email{tomas\_neuman001@ehu.eus}
\affiliation{Centro de F\'{\i}sica de Materiales de San Sebasti\'an, CFM - MPC (CSIC-UPV/EHU), Paseo Manuel Lardizabal 5, 20018 Donostia-San Sebasti\'an, Spain}

\author{Javier Aizpurua}
\email{aizpurua@ehu.eus}
\affiliation{Centro de F\'{\i}sica de Materiales de San Sebasti\'an, CFM - MPC (CSIC-UPV/EHU), Paseo Manuel Lardizabal 5, 20018 Donostia-San Sebasti\'an, Spain}
\affiliation{Donostia International Physics Center (DIPC), 20018 San Sebasti\'an-Donostia, Spain}

\begin{abstract}
We apply the theory of open-quantum systems to describe light emission from coherently driven molecular polaritons. Based on the microscopic Hamiltonian that commonly describes the pure dephasing of isolated molecules, we show that under strong-coupling conditions dephasing leads to a transfer of energy between the constituted polariton branches $|\pm\rangle$. When the polariton dephasing is properly accounted for, the transition from the upper to the lower polariton branch is favored and leads to a dominant population of the lower polariton branch under coherent pumping conditions. As a result, the inelastic light emission originates mainly from the lower polariton state regardless of the pumping laser frequency thus producing an asymmetric emission of light. Furthermore, we show that, when several molecules are considered, inter-molecular coupling breaks the symmetry of the system, making the originally dark polaritons to effectively interact with light.
This effect is revealed in the fluorescence spectrum as new emission peaks. 
\end{abstract}


\maketitle

\section{Introduction}


The interaction of light with molecules has attained much attention due to its potential in photochemical reactivity of molecules \cite{klessinger1995excited}, generation of non-classical states of light \cite{kuhn2006enhancement, akselrod2014probing}, molecular spectroscopy \cite{aroca2013plenhspec, itoh2017reviewempl}, chemical fingerprinting, or in fundamental investigation of single-molecule properties \cite{lidzey1998strong, kuhn2006enhancement, Schull2017, zhang2017electrically, zhang2017sub, imada2017fano, benz2016single, wrigge2008efficient}. This interaction can be enhanced when the molecule is placed into an optical cavity. To that end a large variety of optical cavities have been developed ranging from Fabry-Perot resonators of macroscopic sizes to nanoscale plasmonic cavities where squeezed light is concentrated at (sub)nanometric scales and can interact efficiently with fundamental excitations (e.g. molecular excitons or vibrations) of only a few molecules.  

When placed into optical (plasmonic) cavities, excitons in organic molecules can strongly interact with the cavity modes and form new mixed exciton-photon (plasmon) excitations, so-called exciton-polaritons \cite{lidzey1998strong, hobson2002strongcoupling, bellessa2004strongcoupling, ebbesen2004strongcoupling, michetti2005polstatesdisorder, trugler2008strong, schwartz2011switchingsc, coles2011vibrationally, agranovich2011hybrid, salomon2012scmolarray, kenacohen2013ultrastrongexcpol, vasa2013real, schlater2013plexcitonic, zengin2013approaching, Delga2014, antosiewitz2014enhabstosc,  gulis2015realizing, torma2015strongcrew, george2015ultra, galego2015cavity, galego2016suppressing, zengin2016conditionssc, melnikau2016splittinginlumin, cvik2016scorgpol, herrera2017absphotolumin, Saez-Blazquez2017enhancingcorr, herrera2017dark,  baranov2018novelstrongcoupling, herrera2018review, zeb2018exactvibdressed, Greffet2018nonequilibriumemission}. Exciton-polaritons have been broadly analyzed in connection with their fluorescence properties, cavity-induced (photo)chemistry \cite{hutchinson2012landscapes, canguierdurant2013thermodynamics, galego2015cavity, anoop2016vibreactivity, galego2016suppressing, herrera2016chemistry, flick2017ab}, polariton lasing and polariton condensation \cite{deng2002condensation, kasprzak2006bose, balili2007bose, tosi2012sculpting, plumhof2014room, daskalakis2014nonlinear, sanvitto2016road, dietriche2016pollaser, sun2017PRLboseeinstein} and polariton-mediated energy transfer \cite{coles2014polariton, zhong2017entangled}. 

The inelastic photon emission from the polariton modes has been found to exhibit spectral asymmetries that favor the emission from the lower polariton branch, while often suppresing the emission from the upper polariton \cite{lidzey1999polemission, hobson2002strongcoupling, christogiannis2013characterizing, george2015ultra, wersall2017strongcplpl}. This asymmetry has been attributed to vibrationally driven decay processes between the polaritonic states \cite{Canaguier-Durand2015poldynamics}. It has been shown that the vibrational states of the molecules play a key role in the formation of new vibron-polariton states that lead to the appearance of new peaks in the emission spectra \cite{mukamel1978nature, cvik2016scorgpol, herrera2017absphotolumin, herrera2017dark, herrera2018review, zeb2018exactvibdressed, delpino2018tensor}. Nevertheless, the excitons in organic molecules are also exposed to interactions with their local environment (the solvent) that produces the exciton dephasing. The interaction with the solvent molecules also contributes to significant solvent-dependent photoluminescence Stokes shift induced by the reorganization of the solvent molecules when the solute molecule changes the electronic state \cite{li1994overdampedbrownian, tomasi2005reviewsolvation}. It is therefore necessary to correctly treat the interaction of the polariton states with the dephasing reservoir when describing the strong coupling between the cavity mode and the molecular excitons.  

In this paper we address the inelastic light emission spectra of polaritonic systems pumped by a coherent monochromatic laser. We present a quantum-optical model based on the solution of the quantum master equation \cite{Breuer2005} that describes the spectral asymmetries observed experimentally in the polariton emission and action (excitation) spectra \cite{george2015ultra, wersall2017strongcplpl, hobson2002strongcoupling}. We show that the dominant emission from the lower polariton state is a consequence of the interaction between the excitons and the dephasing reservoir, which in principle includes both the effects of the internal molecular vibrations and the solvent. 

\section{Open quantum system theory of (collective) exciton-cavity-mode coupling}

We describe the molecules as two-level electronic systems composed of the ground, $\lvert {\rm g}\rangle$, and excited state, $\lvert {\rm e}\rangle $, interacting with their respective reservoir, including both the internal molecular vibrational modes and the fluctuations of the local environment of each molecule \cite{tomasi2005reviewsolvation}. The local environment of the molecule is responsible for the electronic dephasing processes [e.g. vibrations of the molecule or the environment \cite{mukamel1978nature, yang1995solventspecden, roy2011phononbathqd, leonarde2014spectraldensities, Canaguier-Durand2015poldynamics, Roy-Choudhury2015phononbath}, fluctuations of solvent polarization etc., as schematically represented in Fig.\,\ref{fig:ch3:polaritondecay1} (a)]. 
The excitonic term of the Hamiltonian of the $i-$th molecule is 
\begin{align}
H_{{\rm e}, i}=\hbar\omega_0\sigma_i^\dagger\sigma_i,
\end{align}
where $\sigma_i$ is the two-level-system lowering operator between the excited state, $\lvert {\rm e}_i\rangle$, and the ground state, $\lvert {\rm g}_i\rangle$, of the $i-$th molecule, $\sigma_i=\lvert {\rm g}_i\rangle\langle {\rm e}_i\rvert$. Each molecule interacts with its local dephasing reservoir described by the Hamiltonian
\begin{align}
H_{\rm res, i}&=\hbar\Omega_{\rm R} B_i^\dagger B_i,\label{eq:ch3:bathham}
\end{align}
via the exciton-reservoir interaction Hamiltonian
\begin{align}
H_{\rm e-res, i}&=d_{\rm R}\Omega_{\rm R}\sigma_i^\dagger\sigma_i (B_i^\dagger +B_i).\label{eq:ch3:bathhamint}
\end{align} 
Here $B_{ i}$ are the bosonic annihilation operators of the collective reservoir mode \cite{li1994overdampedbrownian, toutounji2002brownian, roden2012molsystemsvib, kreisbeck2012lightharvesting, kell2013phononspecden, leonarde2014spectraldensities}, and $\dagger$ stands for the Hermitian conjugate. We have assumed that the reservoir modes have the same frequency $\Omega_{{\rm R,e},i}=\Omega_{{\rm R,g},i}=\Omega_{{\rm R}}$ in the excited state ($\Omega_{{\rm R,e},i}$) and the ground state ($\Omega_{{\rm R,g},i}$). The equilibrium position of the reservoir mode is rigidly displaced in the electronic excited state of the $i-$th molecule by a dimensionless constant $d_{\rm R}$ with respect to its equilibrium position in the ground electronic state. 

The inter-molecular excitonic interactions are assumed to be described through the Hamiltonian
\begin{align}
H_{\rm e-e}=\sum_{ij} G_{ij}\sigma_{i}^\dagger\sigma_{j}+{\rm H.c.},\label{eq:molmol}
\end{align}
where $G_{ij}$ are coupling constants that generally depend on the spatial distribution of the individual molecules as well as on their mutual orientation. 

The molecular excitons interact with a single bosonic cavity mode of frequency $\omega_{\rm c}$ 
\begin{align}
H_{\rm c}=\hbar\omega_{\rm c} a^\dagger a,
\end{align}
where $a$ ($a^\dagger$) is the bosonic annihilation (creation) operator of the cavity mode. The $i$-th molecule interacts with the cavity mode via the coupling Hamiltonian
\begin{align}
H_{{\rm e-c}, i}= \hbar g_i \sigma_i^\dagger a + {\rm H.c.},
\end{align}
where $g_i$ is the respective cavity-mode-exciton coupling constant. 
The total Hamiltonian describing the cavity and molecular excitations finally becomes
\begin{align}
H_{\rm tot}&=H_{\rm c}+H_{\rm e-e}+\sum_i \left(H_{{\rm e}, i}+H_{\rm res, i}+H_{\rm e-res, i}+ H_{{\rm e-c},i} \right).
\end{align}
The Hamiltonian $H_{\rm tot}$ contains the information about the coherent dynamics of the system. However, both the molecule and the cavity are interacting with an environment that causes an effective incoherent decay of their respective excitations. To properly account for this, we describe the dynamics of the system via the master equation for the density matrix $\rho$, including the Lindblad terms of the form $\mathcal{L}_{c_i}(\rho)=\frac{\gamma_{c_i}}{2} \left(2 c_i\rho c_i^\dagger-\lbrace c_i^\dagger c_i,\rho \rbrace \right)$, with $c_i$ the operator of the respective excitation, the phenomenological damping constants of the respective excitations $\gamma_{c_i}$, and with $\lbrace\cdot,\cdot\rbrace$ the anticommutator. 
The quantum master equation that includes all the necessary Hamiltonian and Lindblad terms becomes:
\begin{align}
\dot{\rho}=\frac{1}{{\rm i}\hbar}\left[ H_{\rm tot}, \rho  \right]+\sum_i\mathcal{L}_{c_{i}}(\rho),\label{eq:qme}
\end{align}
where $c_i$ depends on the model under consideration. 
As we detail in the following, the dynamics encompassed in Eq.\,\eqref{eq:qme} leads to the asymmetries observed in the optical response of the strongly-coupled system.

\section{Strong coupling of a single-molecule exciton with a cavity mode}

In the strong coupling regime the plasmon-exciton interaction $g$ becomes so significant that it overcomes the intrinsic electronic ($\gamma_{\sigma_i}$) and cavity ($\gamma_a$) decay rates and leads to the formation of new hybrid states, polaritonic states. The simplest situation arises when a single cavity mode couples with a single two-level electronic system in the single excitation manifold, where only the bare states $\lvert {\rm g},0\rangle$, $\lvert{\rm e},0\rangle$ and $\lvert{\rm g},1\rangle$ are considered, with 0 (1) the number of cavity excitations. The new polaritonic eigenstates $\lvert +\rangle$ and $\lvert -\rangle$ become a coherent admixture of the exciton and the cavity excitation depending on the magnitude of the coupling strength and detuning of their respective frequencies: 
\begin{align}
\begin{split}
\lvert+\rangle &=\cos \theta\lvert{\rm e},0\rangle+\sin \theta\lvert{\rm g},1\rangle,\\
\lvert-\rangle &=-\sin \theta\lvert{\rm e},0\rangle+\cos \theta\lvert{\rm g},1\rangle,
\end{split}\label{eq:transrules}\\
\tan (2\theta)&=\frac{2g}{\omega_0-\omega_{\rm p}}\text{ and } 0<2\theta <\pi.\label{eq:theta}
\end{align}
\begin{figure*}[ht]
\begin{center}
\includegraphics[scale=0.71]{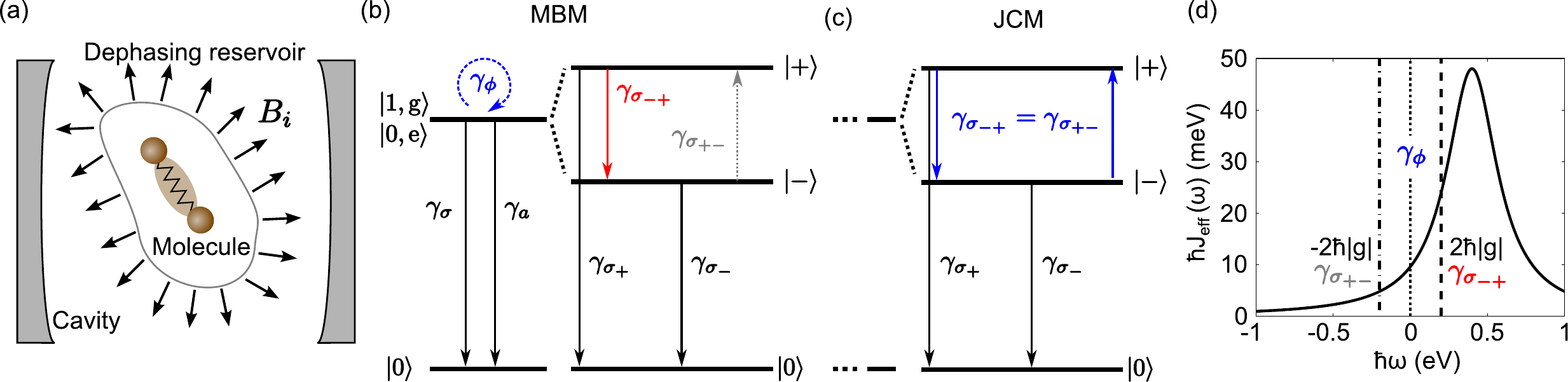}
\end{center}
\caption{Role of the dephasing processes on the light emission from a cavity mode strongly coupled with a single exciton. (a) Schematic representation of the molecule interacting with its dephasing bath containing internal molecular vibrations but also environmental degrees of freedom such as fluctuating polarization of the solvent molecules. The bath modes are represented by bosonic annihilation operators $B_{i}$. (b) Schematic level diagram of the exciton in a cavity that is decoupled (left) and after the coupling is turned on (right) within the Markovian-bath model (MBM). The cavity-exciton coupling gives rise to new polariton states, $\lvert +\rangle$ and $\lvert -\rangle$, and opens new incoherent decay paths between $\lvert +\rangle$ and $\lvert -\rangle$ with the respective rates $\gamma_{\sigma_{-+}}>\gamma_{\sigma_{+-}}$.  (c) The level diagram marking the incoherent population transfer between the polariton states as in (b), but for the Jaynes-Cummings model (JCM) where the rates $\gamma_{\sigma_{-+}}$ and $\gamma_{\sigma_{+-}}$ are equal ($\gamma_{\sigma_{-+}}=\gamma_{\sigma_{+-}}=\sin^2\theta\cos^2\theta\gamma_\phi$). (d) The bath spectral density $J(\omega)$ given by Eq.\,\eqref{eq:specdenlor} for parameters $\hbar\gamma_{\rm R}=400$ meV, $\hbar\Omega_{\rm R}=400$ meV and $d_{\rm R}=0.173$ (for which $\hbar J(0)\approx 20$\,meV). Calculations of selected emission and absorption spectra for smaller values of $\Omega_{\rm R}$ are shown in Supplement 1. The vertical lines mark the positions where the spectral density is evaluated to obtain the values of the Markovian decay rates $\gamma_{\sigma_{\pm\mp}}$ and $\gamma_{\phi}$.}
\label{fig:ch3:polaritondecay1}
\end{figure*}
The scheme of the newly arising energy level structure is schematically drawn in Fig.\,\ref{fig:ch3:polaritondecay1} (b,c). The operators of the three-level system consisting originally of the states $|0\rangle=\lvert{\rm g},0\rangle$, $|2\rangle=\lvert{\rm e},0\rangle$ and $\lvert 3\rangle=\lvert{\rm g},1\rangle$ can be more conveniently expressed in the new basis $\{\lvert 0\rangle, \lvert +\rangle, \lvert -\rangle\}$ with help of Eq.\,\eqref{eq:transrules}. Most importantly, the operator $\sigma^\dagger\sigma$ responsible for the interaction with the dephasing reservoir in $H_{{\rm e-vib}}$ becomes
\begin{align}
\sigma^\dagger\sigma&\approx\lvert 2\rangle\langle 2\lvert =\cos^2\theta \lvert +\rangle\langle+\lvert+\sin^2\theta \lvert-\rangle\langle-\rvert\nonumber\\
&-\sin\theta\cos\theta \left(\lvert-\rangle\langle+\rvert+\lvert+\rangle\langle-\rvert\right).
\end{align}
We further introduce the simplifying notation $\sigma_{ab}=|a\rangle\langle b|$ and rewrite the electron-vibration coupling Hamiltonian as:
\begin{align}
&H_{\rm e-res}=\hbar d_{\rm R}\Omega_{\rm R} \sigma^\dagger\sigma (B^\dagger+B)=\hbar\sigma^\dagger\sigma F\nonumber\\
&=\hbar\left[\cos^2\theta \sigma_{++}+\sin^2\theta \sigma_{--}-\sin\theta\cos\theta \left(\sigma_{-+}+\sigma_{+-}\right)\right]F,
\end{align}
where we have defined $F=d_{\rm R}\Omega_{\rm R}(B^\dagger+B)$. 

Following the standard procedure \cite{Breuer2005}, we now eliminate the dephasing reservoir and derive the incoherent dynamics of the strongly coupled system. To that end we notice that the operators $\sigma_{\pm\mp}$ and $\sigma_{\pm\pm}$ are eigenoperators of the polaritonic Hamiltonian $H_{\rm pol}=H_{\rm c}+H_{{\rm e}}+H_{{\rm e-c}}$ (eigenoperator $O$ of Hamiltonian $H_{\rm pol}$ defined as $[H_{\rm pol},O]=\lambda O$ with $\lambda$ a complex number) and in the interaction picture of $H_{\rm pol}$, these operators have the following time dependence: 
\begin{align}
\sigma_{+-}&=\sigma_{+-}^{(0)}e^{-{\rm i}(\omega_+-\omega_-)t},\\
\sigma_{-+}&=\sigma_{-+}^{(0)}e^{-{\rm i}(\omega_--\omega_+)t},\\
\sigma_{++}&=\sigma_{++}^{(0)}, \\
\sigma_{--}&=\sigma_{--}^{(0)},
\end{align}
with 
\begin{align}
\omega_\pm=\frac{\omega_0+\omega_{\rm c}}{2}\pm\sqrt{g^2+\frac{(\omega_0-\omega_{\rm c})^2}{4}}
\end{align}
the frequency of the upper ($\omega_+$) and lower ($\omega_-$) polariton, respectively.

In the secular approximation, the incoherent processes are represented by the Lindblad terms describing the dephasing of the polariton states, $\mathcal{L}_{\sigma_{++--}}(\rho)$, the decay of $\lvert +\rangle$ to $\lvert -\rangle$, $\mathcal{L}_{\sigma_{-+}}(\rho)$, and the reverse process, $\mathcal{L}_{\sigma_{+-}}(\rho)$.
For brevity we have defined $\sigma_{++--}=\cos^2\theta \sigma_{++}+\sin^2\theta \sigma_{--}$. The respective dephasing and decay rates, $\gamma_\phi=\gamma_{\sigma_{++--}}$, $\gamma_{\sigma_{-+}}$ and $\gamma_{\sigma_{+-}}$, are determined from the properties of the dephasing reservoir characterized by its spectral density $J(\omega)$:
\begin{align}
\gamma_{\sigma_{-+}} &=\cos^2\theta\sin^2\theta {J(\omega_+-\omega_-)},\\
\gamma_{\sigma_{+-}} &=\cos^2\theta\sin^2\theta {J(\omega_--\omega_+)},\\
\gamma_\phi &= {J(0)}.
\end{align}

\begin{figure*}[ht!]
\begin{center}
\includegraphics[scale=0.7]{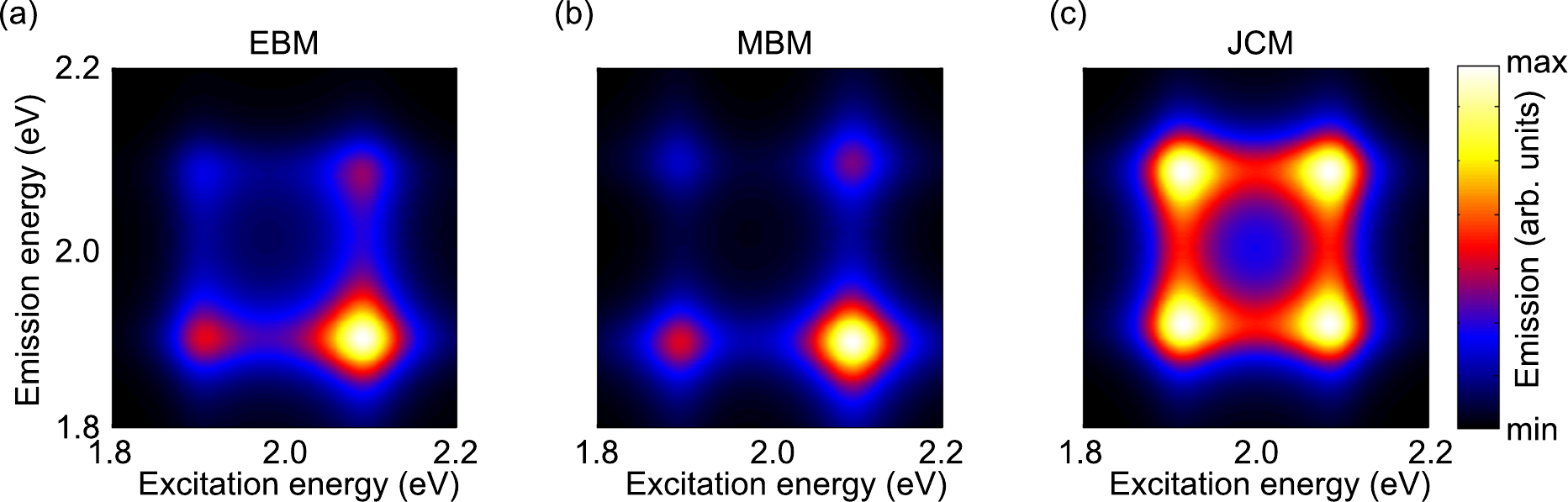}
\end{center}
\caption{Emission spectra normalized to the incident laser intensity $|\mathcal{E}|^2$ as a function of the excitation frequency $\omega_{\rm L}$ within (a) explicit-bath model (EBM), (b) markovian-bath model (MBM) and (c) the Jaynes-Cummings model (JCM). In all calculations we have considered the parameters $\hbar\omega_0=\hbar\omega_{\rm c}=2$ eV, $\hbar\gamma_{\rm a}=150$ meV, $\hbar\gamma_\sigma=2\times 10^{-2}$ meV and $\hbar g=100$ meV. The pure dephasing constant for the JCM is $\gamma_\phi=J(0)$. The parameters of the bath are: $\hbar\gamma_{B}=400$ meV, $\hbar\Omega_{\rm R}=400$ meV and $d_{\rm R}=0.173$.}
\label{fig:ch3:polaritondecay2}
\end{figure*}

The spectral density of the reservoir \cite{kell2013phononspecden, roden2012molsystemsvib, li1994overdampedbrownian, kreisbeck2012lightharvesting, leonarde2014spectraldensities} is obtained as the Fourier transform of the reservoir's two-time correlation function $\langle F^\dagger (t+s) F(t)\rangle$ \cite{Breuer2005}, 
\begin{align}
J(\omega)=2\Re\left\lbrace\int_0^\infty {\rm d} s e^{{\rm i}\omega s} \langle F^\dagger (t+s) F(t)\rangle\right\rbrace,
\end{align}
with $\Re\{\cdot\}$ the real part.
In particular, $J(\omega)$ emerging from Eq.\,\eqref{eq:ch3:bathham} and Eq.\,\eqref{eq:ch3:bathhamint} together with the Lindblad term $\mathcal{L}_{B_i}(\rho)$ (damped harmonic-oscillator reservoir \cite{li1994overdampedbrownian, kreisbeck2012lightharvesting}) calculated for zero temperature, $T=0$\,K, is
\begin{align}
J(\omega)=\frac{2\gamma_{B}d_{\rm R}^2\Omega_{\rm R}^2}{(\Omega_{\rm R}-\omega)^2+\gamma_{B}^2}.\label{eq:specdenlor}
\end{align}
The spectral density $J(\omega)$ of the considered vibrational bath is shown in Fig.\,\ref{fig:ch3:polaritondecay1} (d). $J(\omega)$ has the form of a broad Lorentzian peak positioned at the positive side of the frequency axis. The model parameters used in our study are given in the caption of Fig.\,\ref{fig:ch3:polaritondecay1}. As $J(\omega)$ is not symmetrical with respect to the zero frequency, the transition $|+\rangle\to |-\rangle$ given by the rate $\gamma_{\sigma_{-+}}=\cos^2\theta\sin^2\theta J(2|g|)$ is therefore favored compared to the $|-\rangle\rightarrow |+\rangle$ transition occuring with rate $\gamma_{\sigma_{+-}}=\cos^2\theta\sin^2\theta J(-2|g|)$, [marked by the vertical lines in Fig.\,\ref{fig:ch3:polaritondecay1}(d)]. We stress that this asymmetry is a general property of dephasing reservoirs and robustly appears in a wide range of non-Markovian dephasing models \cite{kell2013phononspecden, roden2012molsystemsvib, toutounji2002brownian, leonarde2014spectraldensities}.This imbalance of the transfer of energy between the polariton states gives rise to the asymmetries observed in the emission spectra \cite{george2015ultra, wersall2017strongcplpl, hobson2002strongcoupling} that we address below. 

Last, in strong coupling we employ the polariton Lindblad operators $\mathcal{L}_{\sigma_{+}}(\rho)$ and $\mathcal{L}_{\sigma_{-}}(\rho)$ ($\sigma_{\pm}=\lvert 0 \rangle\langle \pm \rvert$), where the decay rates of the upper, $\gamma_{\sigma_{+}}$, and the lower, $\gamma_{\sigma_{-}}$, polariton are defined as
\begin{align}
\gamma_{\sigma_{+}}=\sin^2\theta\gamma_a,\\
\gamma_{\sigma_{-}}=\cos^2\theta\gamma_a,
\end{align}
which can be derived from the transformation of the cavity Lindblad term in the secular approximation
\begin{align}
\mathcal{L}_a(\rho)\to\mathcal{L}_{\sigma_{+}}(\rho)+\mathcal{L}_{\sigma_{-}}(\rho),\label{eq:linda}
\end{align} 
and $\gamma_a$ is the decay rate of the bare cavity.
We also include the intrinsic molecular losses via $\mathcal{L}_\sigma(\rho)$, considering $\gamma_\sigma\ll \gamma_a$.

\section{Polariton emission spectra under coherent driving conditions}
\subsection{Single molecule in a cavity}

In the following we consider two different approaches to the implementation of the dephasing due to the reservoir. First, we implement explicitly the reservoir defined by $H_{\rm res,i}$, $H_{\rm e-res, i}$ [Eq.\,\eqref{eq:ch3:bathham} and Eq.\,\eqref{eq:ch3:bathhamint}] and $\mathcal{L}_{B}(\rho)$ into the master equation (the explicit bath model - EBM) and solve for the spectral response. In the second approach we eliminate the dephasing reservoir from Eq.\,\eqref{eq:qme}, as described in the previous section, and introduce the effective dephasing and damping terms via the Lindblad superoperators $\mathcal{L}_\sigma(\rho)$, $\mathcal{L}_{\sigma_+}(\rho)$, $\mathcal{L}_{\sigma_-}(\rho)$, $\mathcal{L}_{\sigma_{++--}}(\rho)$, $\mathcal{L}_{\sigma_{-+}}(\rho)$ and $\mathcal{L}_{\sigma_{+-}}(\rho)$ (the Markovian bath model - MBM). The effective rates are schematically depicted in Fig.\,\ref{fig:ch3:polaritondecay1}\,(b). As a third approach we consider the commonly adopted Jaynes-Cummings model (JCM) where the effective dephasing and decay rates are first defined for the exciton of the molecule and the bare cavity mode, which are mutually decoupled. Note that this is in contrast with the MBM where the incoherent dynamics is derived in the polariton basis. The decay of the cavity and the molecular exciton are described in the JCM by $\mathcal{L}_a(\rho)$ and $\mathcal{L}_\sigma(\rho)$, as defined earlier, and the pure dephasing is implemented via
\begin{align}
\mathcal{L}_{\sigma^\dagger\sigma} (\rho)=\frac{\gamma_\phi}{2}\left( 2\sigma^\dagger\sigma \rho \sigma^\dagger\sigma - \lbrace \sigma^\dagger\sigma, \rho \rbrace \right).
\end{align} 
In the JCM, like in the MBM, the interaction with the reservoir given in Eq.\,\eqref{eq:ch3:bathham} and Eq.\,\eqref{eq:ch3:bathhamint} is not considered. Upon transformation into the polariton basis, the dephasing term in the JCM model yields (among others) interaction terms between $\lvert +\rangle$ and $\lvert -\rangle$, with equal rates for the $\lvert +\rangle\to\lvert -\rangle$ and $\lvert -\rangle\to\lvert +\rangle$ transitions, as schematically depicted in Fig.\,\ref{fig:ch3:polaritondecay1}\,(c). 

As we are interested in the response of the system under illumination by a monochromatic laser light, we introduce the driving term
\begin{align}
H_{\rm pump}=\mathcal{E}\left(a^\dagger e^{-{\rm i}\omega_{\rm L}t}+a e^{{\rm i}\omega_{\rm L}t}\right),\label{eq:ch3:hampump}
\end{align}
with $\mathcal{E}$ the amplitude of the laser pumping and $\omega_{\rm L}$ the laser frequency. We make sure that the pumping amplitude is small enough to conform with the single-excitation approximation. 

We calculate the absorption spectra $s_{\rm A}(\omega)$ of the system (assuming that only the cavity interacts with the radiation field) and the inelastic emission spectra $s_{\rm E}(\omega;\omega_{\rm L})$ for different frequencies $\omega_{\rm L}$ of the incident pumping laser. The spectra are calculated from the quantum regression theorem as one-sided Fourier transforms of the two-time correlation functions (more details are provided in Supplement 1)
\begin{align}
s_{\rm A}(\omega)&=2\Re\int_0^\infty \langle\langle a (\tau)a{^\dagger}(0) \rangle\rangle_{\rm ss}\,e^{\mathrm{i}\omega \tau}\mathrm{d}\,\tau,\label{eq:ch3:abs}\\
s_{\rm E}(\omega;\omega_{\rm L})&=2\Re\int_0^\infty \langle\langle a{^\dagger} (\tau)a(0) \rangle\rangle_{\rm ss}\,e^{-\mathrm{i}\omega \tau}\mathrm{d}\,\tau,\label{eq:ch3:em}
\end{align}
where the double-angle brackets are defined as $\langle\langle a{^\dagger} (\tau)a(0) \rangle\rangle_{\rm ss}=\langle a{^\dagger} (\tau)a(0) \rangle_{\rm ss}-\lim_{\tau\to\infty}\langle a{^\dagger} (\tau)a(0) \rangle_{\rm ss}$.

The calculated emission spectra for the reservoir spectral density assumed in Fig.\,\ref{fig:ch3:polaritondecay1}\,(d) are shown in Fig.\,\ref{fig:ch3:polaritondecay2} within both EBM and MBM and are compared to the result obtained from the JCM. 
To simplify the discussion, in the following we concentrate on the special case when the energies of the plasmonic and excitonic transition are matched ($\omega_{\rm c}=\omega_0$).
In Fig. \ref{fig:ch3:polaritondecay2}(a-c) we plot the emission spectra of the strongly coupled single-molecule exciton with the cavity mode as a function of the excitation frequency $\omega_{\rm L}$ within (a) EBM, (b) MBM, and (c) the JCM. For both the EBM and the MBM the color maps offer the same qualitative and very similar quantitative result. The inelastic emission arises mainly from the transition of the lower polariton to the ground state and thus leads to a clear dominance of the lower polariton emission peak. Contrarily, the JCM yields a fully symmetrical result independently of the excitation frequency, which contradicts the experimental evidences \cite{hobson2002strongcoupling, george2015ultra, wersall2017strongcplpl}. The implementation of the dephasing in the JCM is thus unable to correctly describe the imbalance in the dephasing-driven population transfer between the polaritonic states.
\begin{figure*}[ht]
\begin{center}
\includegraphics[scale=0.7]{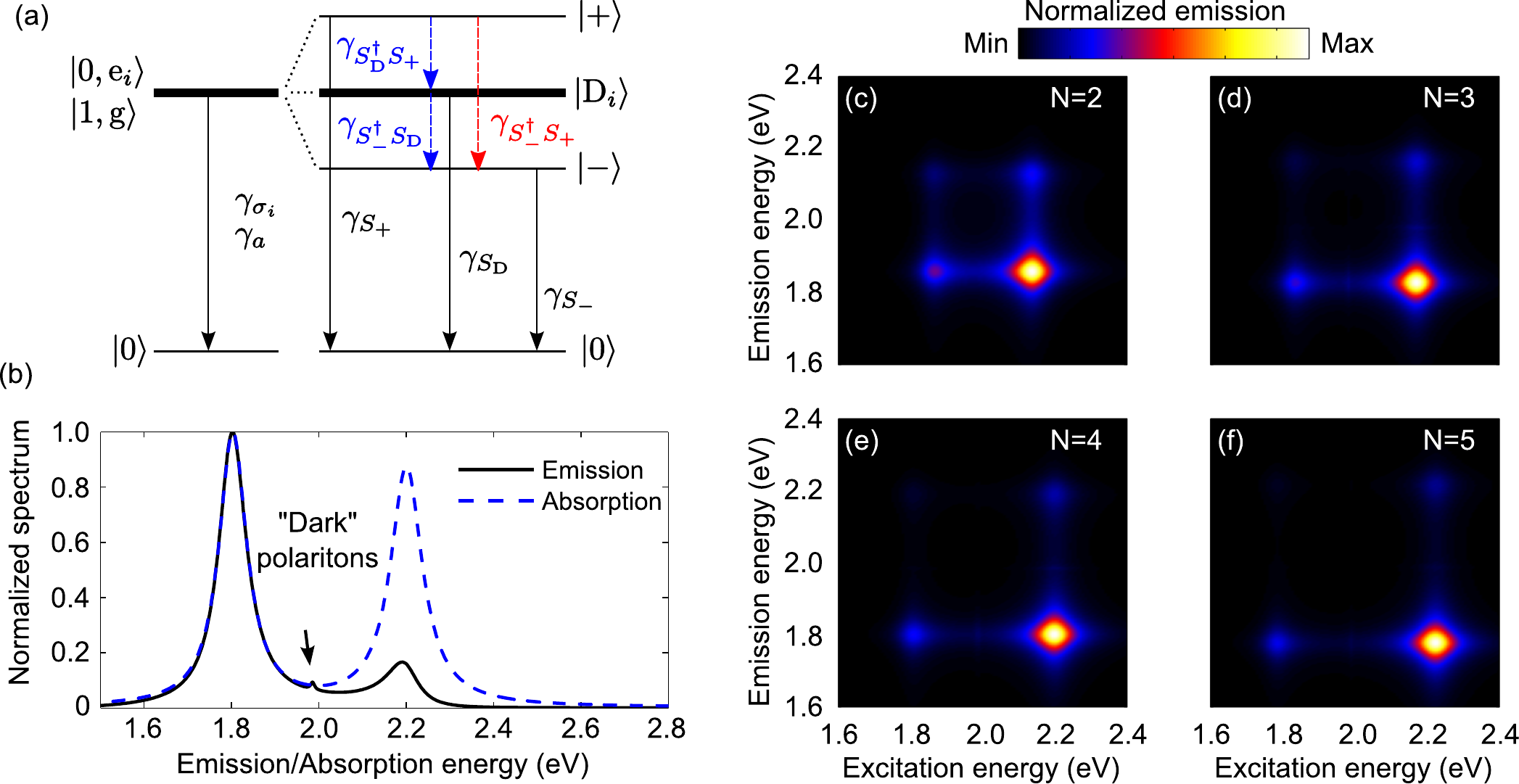}
\end{center}
\caption{(a) Schematic representation of the polariton incoherent dynamics obtained from the full model. The strong coupling leads to formation of bright upper, $\lvert +\rangle$, and lower, $\lvert -\rangle$, polariton that are decoupled from the dark states, $\lvert {\rm D}_i\rangle$. Coupling of the polariton and the dark states with the dephasing reservoir gives rise to the incoherent transfer of populations from the higher-energy states to the lower energy states. The incoherent processes occur with rates $\gamma_a$ for the bare cavity mode, $\gamma_{\sigma_i}=\gamma_{S_{\rm D}}$ for the dark polaritons and the bare molecular excitons, $\gamma_{S_{+}}$ and $\gamma_{S_{-}}$ for the bright polaritons, $\gamma_{S^\dagger_{\rm D}S_+}$, $\gamma_{S^\dagger_-S_{\rm D}}$ and $\gamma_{S^\dagger_-S_{+}}$, respectively. The population transfer is accompanied by the dephasing processes (not shown). (b) Emission (black line) and absorption (blue dashed line) spectra of four molecular excitons ($N=4$) coupled to the cavity mode. The emission from $|-\rangle$ prevails over $|+\rangle$ emission due to the incoherent population transfer caused by the dephasing reservoir. The absorption spectrum, on the other hand, contains both $|\pm\rangle$ peaks of similar intensity. Last, the emission and the absorption spectra contain a peak appearing close to the frequency of the decoupled molecules that arises from the dark polariton states $|D_i\rangle$ that are now coupled to the bright polaritons $|\pm\rangle$. (c-f) Emission spectra as a function of the excitation energy $\hbar\omega_{\rm L}$ for $N=2,\,3,\,4,\,5$ molecules, respectively. In all cases (b-f) the molecular excitons of equal energies $\hbar\omega_0=2$\,eV are perfectly tuned to the cavity resonance $\hbar\omega_{\rm c}=2$\,eV and interact with the cavity mode via $\hbar g_i=\hbar g=100$\,meV. The system is pumped by a laser of amplitude $\hbar\mathcal{E}=0.1$\,meV. The additional parameters are $\hbar\gamma_a=150$\,meV, $\hbar\gamma_{B}=400$\,meV, $\hbar\gamma_\sigma=2\times 10^{-2}$\,meV, $d=0.173$, $\hbar\Omega_{\rm R}=400$\,meV.}\label{fig:absem}
\end{figure*}
\subsection{Many molecules in a cavity}
The strong-coupling between a single-molecule exciton and a cavity mode is fundamentally important but in realistic systems the cavity is usually coupled to several molecular samples \cite{chikkaraddy2016single}. We therefore extend our description to cavities containing $N$ molecules and calculate the absorption and emission spectra as defined in Eq.\,\eqref{eq:ch3:abs} and Eq.\,\eqref{eq:ch3:em} using the EBM. In the EBM we include the Lindblad terms $\mathcal{L}_{B_i}(\rho)$, $\mathcal{L}_{\sigma_i}(\rho)$, $\mathcal{L}_{S_\pm}(\rho)$, where $S_a=\lvert 0\rangle\langle a\rvert$ (with $\lvert 0\rangle$ the ground state and $\lvert\pm\rangle$ the upper (lower) polariton branches). We consider $\gamma_{\sigma_i}=\gamma_\sigma$ and $\gamma_{B_i}=\gamma_B$. More details are provided in Supplement 1. Like in the single-molecular case, the strong coupling of the cavity mode with the excitons of many molecules gives rise to the upper ($\lvert +\rangle$) and lower ($\lvert -\rangle$) polariton branches as schematically depicted in Fig.\,\ref{fig:absem} (a). Additionally to the bright polaritons $\lvert\pm\rangle$, there are $N-1$ states that are decoupled from the cavity (if the intermolecular interaction in Eq.\,\eqref{eq:molmol} preserves the equivalence of all molecules) and are commonly called \textit{dark} polaritons $\lvert {\rm D}_i\rangle$. The polariton states incoherently couple via the dephasing reservoir which allows for population transfer among the bright $\lvert\pm\rangle$ and the dark $\lvert {\rm D}_i\rangle$ polaritons $\gamma_{S^\dagger_{\rm D}S_+}$, $\gamma_{S^\dagger_-S_{\rm D}}$ and $\gamma_{S^\dagger_-S_{+}}$, with decay rates $\gamma_{\sigma_i}=\gamma_{S_{\rm D}}$ for the dark polaritons, $\gamma_{S_{+}}$ and $\gamma_{S_{-}}$ for the bright polaritons, respectively. In our notation, $S_a=\lvert 0\rangle\langle a\rvert$ (with $\lvert 0\rangle$ the ground state). The newly arising states also undergo dephasing (not shown in the schematics), in analogy with the case of the single exciton. More details about the respective processes are provided in Supplement 1. As opposed to the case where only a single molecule is considered, in the collective scenario the dark polariton states mediate the population decay $\lvert +\rangle\to\lvert -\rangle$ and change the population dynamics observed for the lower polariton state if $\lvert + \rangle$ is pumped.

In the following we assume an inter-molecular coupling of the form
\begin{align}
\begin{split}
G_{ij}&=\frac{G_0}{\lvert i-j\rvert^3}\;{\rm for}\;i\neq j \; {\rm and}\\
G_{ij}&=0\;{\rm for}\;i=j
\end{split}\label{eq:intermol}
\end{align}
and set 
\begin{align}
\hbar G_0=\frac{p_0^2}{4\pi\varepsilon_0 r_0^3},
\end{align}
with the transition dipole moment of the exciton $p_0=0.2$\,{e$\cdot$nm}, the effective intermolecular distance $r_0=2$\,nm and $\varepsilon_0$ the vacuum permittivity. This choice of $G_{ij}$ describes a set of interacting molecules whose dipoles are arranged along a line (e.g. in the $x$ direction) with constant spacing $r_0$ and with parallel dipole moments $p_0$ (e.g. oriented along $z$).
The intermolecular interaction given by Eq.\,\eqref{eq:intermol} weakly perturbs the polariton structure given by the collective cavity-mode-exciton Hamiltonian but breaks the symmetry of the Hamiltonian (makes the molecules inequivalent).  Due to this, the originally dark polariton states $|{\rm D}_i\rangle$ couple with the cavity mode and become observable in the spectra. We discuss more details of the collective-coupling model in Supplement\,1.

As an illustrative example we calculate the emission and absorption spectra of four mutually interacting molecules that are coupled to the cavity with $\hbar g_i=\hbar g=100$\,meV. The system is pumped at the upper polariton frequency $\hbar\omega_{\rm L}=2.2$\,eV. The result is shown in Fig.\,\ref{fig:absem} (b) for $N=4$ molecules interacting with the cavity mode. The emission spectrum (black solid line) shows a dominant peak originating from the lower polariton $|-\rangle$ (appearing at $\approx 1.8$\,eV) as in the single-molecular case. Another sharp emission peak of low intensity, which was not present in the single-molecular case, emerges at a frequency around that of the decoupled molecules $\approx 2$\,eV. This new peak is a signature of the polariton states $|D_i\rangle$ that are dark in the collective-coupling model where the excitons do not interact directly among themselves, but become bright after introducing the inter-molecular coupling in Eq.\,\eqref{eq:molmol}. Experiments where large numbers of molecules couple with the cavity show that the dark-polariton photoluminescence peak can have comparable intensity to the emission peak of the lower polaritons \cite{george2015ultra, wersall2017strongcplpl}. On the other hand, the absorption spectrum (blue dashed line) features two absorption peaks of commensurate intensity at frequencies of the $|\pm\rangle$ polariton branches. As a result of the inter-polariton transfer induced by the reservoir, the lower-polariton peak has slightly higher spectral intensity and is narrower than the upper-polariton peak. 

Finally, in Fig.\,\ref{fig:absem} we present the two dimensional maps containing the emission (vertical axis) and excitation (horizontal axis) spectra of systems containing (c) $N=2$, (d) $N=3$, (e) $N=4$ and (f) $N=5$ molecules (considering $\hbar g=100$\,meV). The emission pattern is in all the cases similar to the single-molecule case [Fig.\,\ref{fig:ch3:polaritondecay1}(d,e)], exhibiting a doublet of the emission peaks originating from $|\pm\rangle$ that are split by the collectively enhanced coupling $\sqrt{N}g$. 
Between the $\lvert +\rangle$ and $\lvert -\rangle$ polariton peaks, in this collective scenario there appears and additional feature corresponding to the dark polaritons in both the emission and the excitation spectra, which is hardly distinguishable in the spectral maps.
The dominance of the lower-polariton peak in all calculated spectra is in accordance with the mechanism of incoherent population transfer in strongly-coupled systems discussed above. We can observe that the inelastic emission is most efficient from the lower polariton branch when the upper polariton is pumped. In this case, the interaction with the reservoir efficiently incoherently populates $\lvert -\rangle$ which in turn emits the inelastic photons. We now briefly analyze the polariton dynamics in the collective scenario that gives rise to the asymmetry of the inelastic photon emission.

\section{Polariton dynamics in the collective scenario}

We have shown that the dephasing reservoir gives rise to incoherent transitions between the polariton states that preferentially lead from the states of higher energy towards the states of lower energy ($\lvert +\rangle\to\lvert {\rm D}_{i}\rangle$, $\lvert +\rangle\to\lvert -\rangle$ and $\lvert {\rm D}_{i}\rangle\to\lvert -\rangle$). Here we focus on the dynamics of these decay processes and calculate the time evolution of the polariton populations $n_+=\langle S^\dagger_+S_+ \rangle$, $n_-=\langle S^\dagger_-S_- \rangle$ and $n_{\rm D}=\frac{1}{N-1}\sum_i\langle S^\dagger_{{\rm D}_i}S_{{\rm D}_i} \rangle$ assuming that the populations evolve according to the master equation [Eq.\,\eqref{eq:qme} with Eq.\,\eqref{eq:linda}] that explicitly includes the dephasing reservoir (the EBM). We compare the EBM population dynamics with a rate-equation model (REM) based on the diagram of levels and decays displayed in Fig.\,\ref{fig:absem}\,(a) (more details are provided in Supplement 1). We calculate the dynamics assuming that the upper polariton is initially fully populated $n_+=1$  and $n_-=n_{\rm D}=0$ and then spontaneously decays (the coherent driving [Eq.\,{eq:ch3:hampump}] is switched off) into the ground state $\lvert 0\rangle$ and into the other polariton states, $\lvert-\rangle$ and $\lvert {\rm D}_i\rangle$. 
\begin{figure}[h!]
\begin{center}
\includegraphics[scale=0.9]{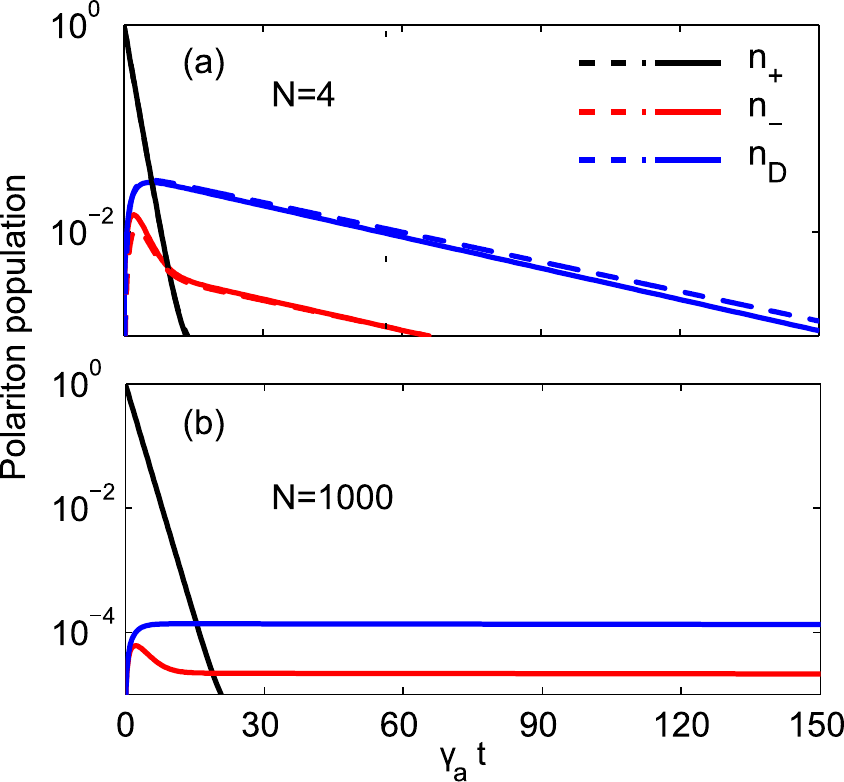}
\end{center}
\caption{Decay of polariton populations $n_+$ (upper polariton - black), $n_-$ (lower polariton - red) and $n_{\rm D}$ (dark polariton - blue), in logarithmic scale, as a function of time assuming that initially $n_+=1$ and $n_-=n_{\rm D}=0$. (a) The full calculation (EBM - dashed lines) is compared to the rate-equation model (REM - full lines) for $N=4$ molecules and $g=100$\,meV. (b) The populations calculated from the REM for $N=1000$ molecules, using $\sqrt{N}g=200$\,meV. The remaining parameters are $\hbar\gamma_a=150$\,meV, $\hbar\gamma_{B}=400$\,meV, $\hbar\gamma_\sigma=2\times 10^{-2}$\,meV, $d=0.173$, $\hbar\Omega_{\rm R}=400$\,meV. }\label{fig:ch3:polaritondecay3}
\end{figure}

In Fig.\,\ref{fig:ch3:polaritondecay3} we plot the polariton populations in the logarithmic scale as a function of time obtained from the numerical time evolution of the full system-reservoir density matrix (EBM - dashed lines) together with the solution of the REM (full lines) for (a) $N=4$ molecules and (b) $N=1000$ molecules (using REM only). For $N=4$, the REM matches well with the EBM with only slight deviations from the exact population dynamics. The $n_+$ (black) exhibits a rapid decay [$\gamma_{S_+}+\gamma_{S^\dagger_- S_+}+(N-1)\gamma_{S^\dagger_{\rm D} S_{+}}$] from its original population into the ground state ($\gamma_{S_+}$), but also into the lower polariton, $\lvert -\rangle$ ($\gamma_{S^\dagger_- S_+}$), and the dark polaritons, $\lvert {\rm D}_i\rangle$ [$(N-1)\gamma_{S^\dagger_{\rm D} S_{+}}$], which pumps the lower ($n_-$ - red) and dark polariton ($n_{\rm D}$ - blue) populations, respectively. After this initial impulse, the dark polariton population starts to steadily decay into the ground state ($\gamma_{S_{\rm D}}$) and into the lower-polariton state ($\gamma_{S^\dagger_{\rm -} S_{\rm D}}$). Similarly to the dark polaritons, the lower polariton first gets populated due to the fast-decaying upper polariton. After that $\lvert -\rangle$ rapidly decays into the ground state ($\gamma_{S_{-}}$), but only until it reaches the regime when $n_-$ is dominantly pumped by the slowly decaying dark-polariton ($\gamma_{S^\dagger_{-}S_{\rm D}}$). In this regime, the decay of $n_{-}$ becomes limited by the pumping and resembles that of the dark polaritons (the \textit{bottleneck} effect, red lines in Fig.\,\ref{fig:ch3:polaritondecay3}\,(a)). 

We show in Supplement 1 that the decay rates connecting the polariton states are inversely proportional to the number of molecules $\gamma_{S^\dagger_{b} S_{a}}\propto 1/N$. Since the upper and the lower polaritons in our model decay fast to the ground state ($\gamma_{S_\pm}\propto\gamma_a$) regardless of $N$, the initial stages of their respective population dynamics are practically independent of the number of molecules. However, as $N$ is increased, the dark polariton decay rate into the lower polariton becomes progressively smaller ($\gamma_{S^\dagger_-S_{\rm D}}\propto 1/N$) until it becomes fully limited by the intrinsic rate $\gamma_\sigma$ for $N\to\infty$. This tendency is apparent from Fig.\,\ref{fig:ch3:polaritondecay3}\,(b), where we plot the population decay for $N=1000$ molecules as obtained from the rate-equation model. 

Finally we remark that the model described in this paper is able to address the dynamics of population transfer among the polaritonic states, but does not explain the long lifetime of the lower polariton state that has been reported in the literature \cite{virgili2011dynamics, schwartz2013poldynamics, wang2014quantumyield, Canaguier-Durand2015poldynamics}. In our approach, the terminal slow decay of $n_-$ arises due to the \textit{bottleneck} in the form of a slowly decaying \textit{dark} polariton states. The explanation of the long lower-polariton lifetime requires further modeling of the microscopic decay mechanisms of the coupled cavity mode and the molecular excitons.    

\section{Conclusion}

In conclusion, we have demonstrated that the dephasing reservoir in strongly coupled cavity-mode-exciton systems can lead to asymmetries in the observed emission spectra, favoring the light emission from the lower polariton and suppressing the upper polariton emission. The asymmetry in the inelastic light emission from the cavity arises naturally from the model which explicitly considers the dephasing bath as an effective damped-harmonic oscillator. The coupling with the reservoir in the strong coupling regime naturally favors the transfer of the population of the higher-energy polaritons towards the polaritons of lower energy ($\lvert +\rangle\to\lvert {\rm D}_{i}\rangle$, $\lvert +\rangle\to\lvert -\rangle$ and $\lvert {\rm D}_{i}\rangle\to\lvert -\rangle$), including the dark polaritons if many molecules are considered. This process leads to the prevalence of the inelastic photon emission from the lower polariton $\lvert -\rangle$ and considerably shorter lifetime of the upper polariton $\lvert +\rangle$. Moreover, if many mutually interacting molecules are coupled to the cavity, the dark polariton states can become bright and give rise to a new peak in the polariton emission spectrum. This new peak is then positioned approximately at the frequency of the uncoupled excitons, which is consistent with experimental observations \cite{george2015ultra, wersall2017strongcplpl, hobson2002strongcoupling}.

The results presented in this paper provide an intuitive view of the processes that stand behind the experimental observations and can serve as guidelines for future implementations of dephasing in strongly-coupled systems.
\section{Acknowledgments}
We acknowledge T. Ebbesen for inspiring discussions.
 
\section{Funding}

Project FIS2016-80174-P from the Spanish Ministry of Economy and Competitiveness (MINECO).

\subsection*{}
See Supplement 1 for supporting content.

\providecommand{\noopsort}[1]{}\providecommand{\singleletter}[1]{#1}%
%




\end{document}